\input harvmac

\def\bF{\bar\Phi}
\def\bP{\bar\Psi}

\def\bphi{\bar\phi}

\def\bpsi{\bar\psi}
\def\cA{{\cal A}}
\def\cD{{\cal D}}
\def\cd{C_D}
\def\d{\partial}
\def\Ga{\Gamma}
\def\G11{\Gamma_{11}}
\def\g{\gamma^}
\def\h{{1 \over 2}}

\def\sD{\not\!\! D}

\def\so{$SO(10)\;$}

\Title{\vbox{\baselineskip12pt\hbox{INRNE-TH-99/1}}}
{\vbox{\centerline{On the Generalised Pauli--Villars Regularization}
\vskip2pt\centerline{of the Standard Model}}}

\centerline{Michail Nicolov Stoilov\footnote{*}{
e-mail: mstoilov@inrne.bas.bg}
}

\bigskip\centerline{Institute for Nuclear Research and Nuclear Energy}
\centerline{Boul. Tzarigradsko Chaussee 72, 1784 Sofia, Bulgaria}

\vskip .3in

\noindent{\bf Abstract. }
We show that the regularization
of the Standard Model
proposed by Frolov and Slavnov
describes a nonlocal theory with
quite simple Lagrangian.

\vskip .3in


\Date{22 February 1999}
\vfil\eject
\baselineskip20pt

\nref\base{S.A. Frolov and A.A. Slavnov, {\it Phys. Lett.} {\bf B309}
(1993) 344.}
%
\nref\nn{R. Narayanan and H. Neuberger, {\it Phys. Lett.} {\bf B302}
(1993) 62.}
%
\nref\ak{
S. Aoki and Y. Kikukawa, {\it Mod. Phys. Lett.} {\bf A8} (1993) 3517.}
%
\nref\fu{
K. Fujikawa, {\it Nucl. Phys.} {\bf B428} (1994) 169.}
%
%
The construction of gauge invariant regularization for
the chiral theories (even for the anomaly free ones) was an open problem
for a long time.
Only recently such a regularization was proposed for the Standard Model
\base ;
some generalisations and important insights on the
issue can be found in Refs.\nn --\fu .
The new regularization (called generalised
Pauli--Villars regularization) is a version of the standard
gauge-invariant Pauli--Villars one where one regularizes entire loops,
but not separate propagators.
The main difference is that infinitely many regulator fields are used.
Therefore, in order to specify the regularization completely, one needs
for any divergent diagram a recipe for how to handle the infinite
sum of the terms due to regulator fields.
In this letter our aim is to show that the contribution of the regulator
fields can be calculated on the
Lagrangian level, so as to give a nonlocal theory.

We begin with a short description of the generalised Pauli--Villars
regularization of the Standard Model.
%
\nref\wz{F. Wilczek and A. Zee, {\it Phys. Rev.} {\bf D25} (1982)
553-565.}
In Ref.\base\
a construction \wz\  is used, where all one-generation matter fields
are combined into a single chiral \so spinor $\psi_+$
(which is also a chiral Lorentzian spinor)
and all gauge fields --- into an \so gauge field.
The gauge field Lagrangian is regularized by the higher covariant
derivative method and is not considered in \base\ (and neither is here).
In addition to the original fields an infinite set of commuting and
anticommuting Pauli--Villars fields ($\phi_r$ and $\psi_r$
respectively, $r\ge1$) is added.
These new fields are simultaneously chiral Lorentzian spinors
and non-chiral \so ones.
The explicit form of the mass terms for the regulator fields is
determined by the requirement that they are nonzero, real,
\so and Lorentzian scalars
and by the chirality properties of the fields.
As a result the Lorentz and \so charge conjugation matrices ($\cd$ and
$C$) have to be used.
(Basic feature of any charge conjugation matrix ${\cal C}$
is that ${\cal C}\psi$
transforms under the conjugate representation to that of $\psi$.)
A list of properties of $\cd$ and $C$ used in this work is given
in Appendix.

The one-generation matter field regularized Lagrangian
of the Standard Model reads
\eqn\first{\eqalign{
L_{reg} & = \bpsi_+ i\sD\psi_+ \cr
        & + \bpsi_r i\sD\psi_r
+\h M_r(\psi_r^T\cd C \G11\psi_r+\bpsi_r\cd C\G11\bpsi^T_r) \cr
        & + \bphi_r \G11 i\sD\phi_r
-\h M_r(\phi_r^T\cd C \phi_r - \bphi_r\cd C\bphi^T_r).\cr
}}
Here $\sD$ is the covariant derivative with respect to
the \so gauge field $\cA$,
$\sD = \g{\mu}(\d_\mu -i g \cA_\mu^{ij}\sigma_{ij})$,
$\sigma_{ij}$ are the \so generators;
$M_r = M r$, where $M$ is a (large) mass parameter (Pauli--Villars mass)
and a summation over $r\ge 1$ is assumed.
This form of $M_r$ is crucial for the convergence of the diagrams in the
model \nn\ while the concrete sign of $M_r$ does not matter.
Introducing the projectors on the irreducible spinor representations:
$$
\Pi_\pm=\h(1\pm\g{5})\;\;\; {\rm and}\;\;\; P_\pm=\h(1\pm\G11)
$$
the chirality properties of the fields read:
$$
\eqalign{
\Pi_-\psi_+ &= \Pi_-\psi_r = \Pi_-\phi_r = 0,\cr
P_-\psi_+ &= P_\pm \psi_{r\mp} = P_\pm \phi_{r\mp} = 0,\cr
}
$$
where
$\psi_r=\psi_{r+}+\psi_{r-}$
and analogously for $\phi_r$.

Any \so gauge model is anomaly free, and so it is not a big surprise
that \first\ could be rewritten in a vector-like form.
Following \ak\ we introduce variables
\eqn\newvar{\eqalign{
\Psi_r &= \psi_{r+} + C\cd\bpsi_{r-}^T,\cr
\Phi_r &= \phi_{r+} + C\cd\bphi_{r-}^T. \cr
}}
Both these new fields are \so chiral and Lorentzian non-chiral spinors
contrary to the original ones.
$$
\eqalign{
P_-\Psi_r &= P_-\Phi_r = 0,\cr
\Pi_+\Psi_r = \psi_{r+}; &\;\;\;\Pi_-\Psi_r = C\cd\bpsi_{r-}^T, \cr
\Pi_+\Phi_r = \phi_{r+}; &\;\;\;\Pi_-\Phi_r = C\cd\bphi_{r-}^T. \cr
}
$$
Using definitions \newvar\
and the ones following from them
$\bP_r = \bpsi_{r+} - \psi_{r-}^T C\cd$ and
$\bF_r = \bphi_{r+} - \phi_{r-}^T C\cd$,
eq.\first\ takes the form
\eqn\second{
L_{reg}=\bpsi_+i\sD\psi_+ + \bP_r(i\sD-M_r)\Psi + \bF_r(i\sD+M_r)\Phi.
}
The Berezian corresponding to the change of variables
\newvar\ is $1$, which guarantees
that Lagrangians \first\ and \second\ describes one and the same theory.

Now we want to reformulate \second\ as a higher derivative theory.
%
\nref\my{M.N. Stoilov, {\it Ann. Physik} {\bf 7} (1998) 1}
Following \my\ our first step is to replace the commuting Pauli--Villars
fields by anticommuting ones.
The idea is to consider instead of
\eqn\bose{
L=\bF(i\sD - M)\Phi
}
the following one
\eqn\new{
L=(\bF+\bar\chi)(i\sD - M)(\Phi+\chi),
}
where $\chi$ is an additional dynamical field.
This Lagrangian has a very large Stuckelberg-type gauge symmetry.
Its fixing produces Faddeev--Popov ghosts
$(\eta\;\; {\rm and}\;\;\bar\eta)$
which have statistics, opposite to $\Phi$, i.e.  they are normal
anticommuting spinors.
A particular gauge choice brings \new\ into \bose\ plus
ghosts terms trivially
decoupled from the dynamics;
another gauge choice leaves only
\eqn\fermi{
L'= -\bar\eta(i\sD - M)\eta
}
(plus decoupled $\Phi$-terms).
Thus eqs.\bose\ and \fermi\ describe the same physics.
This is true provided there are no sources for the field $\Phi$ (and $\eta$)
in the model and this is exactly the situation with
the Pauli--Villars fields.
Applying the procedure described above to all $\Phi_r$
the Lagrangian \second\ could be rewritten as:
\eqn\third{
L_{reg}=\bpsi_+i\sD\psi_+ + \bP_r(i\sD+M_r)\Psi - \bar\eta_r(i\sD-M_r)\eta,
}
where all fields are anticommuting now.

Our next step is to combine different terms in \third\ into a single higher
derivative Lagrangian.
It was shown in \my\ that the Lagrangian
$$
L=g\bpsi(i\sD-m_1)(i\sD-m_2)\psi
$$
after suitable Legendre transformation could be put into the form
$$
L={g\over\vert g\vert}\left(
\bpsi_1(i\sD-m_1)\psi_1 - \bpsi_2(i\sD-m_2)\psi_2 \right)
$$
and vice-versa, provided $m_1 > m_2$.
Note that up to a sign the last Lagrangian
is independent of the coupling constant $g$.
Such decomposition into a sum of first order Lagrangians holds
for any higher derivative Lagrangian
$L=\bpsi\prod_i(i\sD-m_i)\psi$, if $m_i\ne m_j\;\;\forall
i,j: i\ne j$.
Here we use these results to bring together different terms in \third .
We combine Pauli--Villars terms with equal $r$ and obtain
$$
L_{reg}=\bpsi_+i\sD\psi_+ - \bP_r(1- {\sD^2\over M^2r^2})\Psi,
$$
where we have used the freedom to the choose coupling constant to be
$g_r=(M r)^{-2}$.
In this way we have introduced a small parameter $(1/M)$ which counts
the order of the derivatives in the Lagrangian.
%
\nref\JLM {X. Ja\'en, J. Llosa, A. Molina,  Phys. Rev.
{\bf D34} (1986) 2302.}
Such a parameter is needed in any higher derivative theory \JLM .
It automatically produces the so called {\it perturbative constraints}
which exclude the unwanted negative norm states (those of Pauli--Villars
fields in our case).
Then we ``add'' one after the other the new second derivative terms to the
matter field term and get
\eqn\almost{
L=\bpsi i\sD
  \prod_{r=1}^\infty\left(1-{\sD^2\over M^2r^2}\right).
}
The solution of the equation of motion following from the Lagrangian
\almost\ is
$\psi = \sum_{-\infty}^\infty \psi_r$
where
$(i\sD-M r)\psi_r = 0$.
It describes the full set of the matter and Pauli--Villars fields.
The only remnant from the perturbative constraints is
\eqn\constr{\Pi_-\vert{\rm physical}> = 0}
which excludes the Pauli--Villars fields from the physical ones.
Finally, using the formula
$$ \sin{(x)}= x\prod_{r=1}^\infty\left(1 - {x^2\over\pi^2r^2}\right)$$
we obtain
\eqn\final{
L_{reg}=M\bpsi\sin{\left({i\sD\over M}\right)}\psi.
}
Eq.\final\ together with the constraint \constr\ give us the nonlocal
version of the
regularized Lagrangian of the Standard Model.

A final note: Lagrangian \final\ could be rewritten as
$L_{reg}=\bP i\sD f(\cD^2/M^2)\Psi$,
where $f(x^2) = \sin{x}/x$.
In this form it is closely related to the interpretation of
the generalised Pauli--Villars regularization discussed in \fu .

\vskip 1cm
\centerline{\bf Appendix}
\vskip 5mm

%
\nref\bs{N.N. Bogoliubov and D.V. Shirkov, {\it Introduction in Quantum
Field Theory}, Moscow, Nauka, {\bf 1973}}
{\bf Lorentz charge conjugation matrix \bs .}

In $4$-dimensional space-time with metrics
$g^{\mu\nu} = diag(1,-1,-1,-1)$
the gamma matrices $\g{\mu}, \mu=0,1,2,3$ are such, that
$(\g{\mu})^\dagger = \gamma_\mu$.
The matrix $\g{5}$  is defined as $\g{5}=-i\g{0}\g{1}\g{2}\g{3}$ and so
$(\g{5})^\dagger=(\g{5})^T=\g{5}$.
The charge conjugation matrix $\cd$ (the one used here
is inverse to that in \bs )
has the properties:
$$
\eqalign{
\cd^{-1}(\g{\mu})^T\cd &=-\g{\mu},\;\;\; \mu=0,\dots, 3\cr
\cd^{-1}\g{5}\cd &=\g{5},\cr
\cd = \bar\cd = -\cd^\dagger & = - \cd^T = - \cd^{-1},\cr
\det{\cd}^2 &= 1.\cr
}
$$

{\bf \so charge conjugation matrix \wz .}

Corresponding gamma matrices $\Ga_i,\;\; i=1,\dots,10$
are $32\times 32$ matrices, such that
$\{\Ga_i,\Ga_j\} = \delta_{ij}$ and
$\Ga_i^\dagger = \Ga_i\;\; \forall i$.
$\G11$ matrix is defined as
$\G11=-i\Ga_1.\dots.\Ga_{10}$, so that
$\G11^\dagger=\G11^T=\G11$.
The generators of the \so algebra in the spinorial representation are
$\sigma_{ij}=\h i [\Ga_i,\Ga_j], \;\; i\ne j,\;\;i,j=1,\dots,10$.
The charge conjugation matrix $C$ has the following properties:
$$
\eqalign{
C^{-1}\Ga_i^T C &=-\Ga_i,\;\;\; i=1,\dots,11\cr
C^{-1}\sigma_{ij}^T C &=-\sigma_{ij},\cr
C = \bar C = -C^\dagger &= - C^T = - C^{-1},\cr
\det{C}^2 &=1.\cr
}
$$
%


\listrefs
\bye

\end